\documentclass[pdflatex,sn-mathphys]{sn-jnl}

\usepackage{longtable} 
\usepackage{pdflscape} 
\usepackage{pdfpages}

\jyear{2021}%

\theoremstyle{thmstyleone}%
\theoremstyle{thmstyletwo}%

\theoremstyle{thmstylethree}%

\begin{document}

\title[Article Title]{Inconsistencies in the Definition and Annotation of Student Engagement in Virtual Learning Datasets: A Critical Review}


\author[1]{\fnm{Shehroz} \sur{S. Khan}}\email{shehroz.khan@uhn.ca}

\author*[1]{\fnm{Ali} \sur{Abedi}}\email{ali.abedi@uhn.ca}

\author[1]{\fnm{Tracey J.F.} \sur{Colella}}\email{tracey.colella@uhn.ca}

\affil[1]{\orgdiv{KITE Research Institute}, \orgname{University Health Network, Canada}}

\abstract{\textbf{Background:} Student engagement (SE) in virtual learning can have a major impact on meeting learning objectives and program dropout risks. Developing Artificial Intelligence (AI) models for automatic SE measurement requires annotated datasets. However, existing SE datasets suffer from inconsistent definitions and annotation protocols mostly unaligned with the definition of SE in educational psychology. This issue could be misleading in developing generalizable AI models and make it hard to compare the performance of these models developed on different datasets. The objective of this critical review was to explore the existing SE datasets and highlight inconsistencies in terms of differing engagement definitions and annotation protocols.
\textbf{Methods:} Several academic databases were searched for publications introducing new SE datasets. The datasets containing students' single- or multi-modal data in online or offline computer-based virtual learning sessions were included. The definition and annotation of SE in the existing datasets were analyzed based on our defined seven dimensions of engagement annotation: sources, data modalities, timing, temporal resolution, level of abstraction, combination, and quantification.
\textbf{Results:} Thirty SE measurement datasets met the inclusion criteria. The reviewed SE datasets used very diverse and inconsistent definitions and annotation protocols. Unexpectedly, very few of the reviewed datasets used existing psychometrically validated scales in their definition of SE.
\textbf{Discussion:} The inconsistent definition and annotation of SE are problematic for research on developing comparable AI models for automatic SE measurement. Some of the existing SE definitions and protocols in settings other than virtual learning that have the potential to be used in virtual learning are introduced.
}

\keywords{Virtual Learning, Student Engagement, Engagement Measurement, Engagement Definition, Artificial Intelligence, Machine Learning}



\maketitle

\section{Introduction}
\label{sec:introduction}
As the use of internet services becomes more widespread, virtual learning programs are becoming increasingly common and accepted as a mainstream form of education \cite{mukhtar2020advantages}. In contrast to traditional in-person learning, virtual learning offers several benefits, including increased accessibility, lower costs, and the ability to provide personalized instruction \cite{dung2020advantages}. However, virtual learning also presents its own set of challenges, particularly when it comes to assessing Student Engagement (SE). In a virtual learning environment, it can be difficult for instructors to measure the level of engagement of their students, especially when working with large groups in online virtual learning settings \cite{sumer2021multimodal}. This is a significant issue because SE has been shown to have a direct impact on the achievement of learning objectives \cite{gray2016effects}. Therefore, it is important for instructors to be able to assess SE in order to provide real-time feedback and take necessary actions to maximize engagement.

In recent years, advances in Artificial Intelligence (AI) have led to the successful development of algorithms to objectively and automatically measure SE in virtual learning environments, especially in academia and online classrooms \cite{karimah2022automatic,dewan2019engagement}. Most of the published results in this area rely on supervised machine-learning approaches \cite{karimah2022automatic,dewan2019engagement}, requiring annotated ground-truth data to develop models and to provide SE-related outcomes (e.g., Engaged versus Not-engaged or different levels of engagement). A major concern is that most of the datasets used non-standard definitions of engagement; thus, the data samples in many of these datasets are annotated very differently across multiple datasets. Unless a standardized SE definition and measurement scale are in place, annotating data to develop AI algorithms is very challenging. This further constrains the development of AI algorithms to objectively quantify SE and to compare the SE measurement algorithms fairly.

The objective of this critical review was to identify inconsistencies in definitions and annotation protocols used in the existing SE datasets. The research question of this study was as follows: How inconsistent was the definition and annotation of SE in the existing datasets based on seven dimensions of engagement annotation (described in Section \ref{sec:dimensions}), including
\begin{enumerate}
        \item Sources: the observers performing the annotation,
        \item Data modality: the information that is observed by the observers for annotation,
        \item Timing: the time when the annotation takes place,
        \item Temporal resolution: the timesteps in which the annotation takes place,
        \item Level of abstraction: whether engagement is defined and annotated as a single- or multi-component variable,
        \item Combination: the way the components of engagement are combined to create one value for engagement, and
        \item Quantification: the way the engagement is represented numerically.
\end{enumerate}

\section{‌Student Engagement Annotation}
\label{sec:student_engagement_annotation}
Researchers have identified SE to encompass three primary components \cite{fredricks2004school,trowler2010student}. These components include behavioral engagement, which refers to behaviors such as attendance, involvement, and being On-Task; affective engagement, which refers to emotional reactions such as excitement and desirability; and cognitive engagement, which refers to a student's investment in learning and willingness to embrace challenges. An additional dimension of SE, referred to as "agentic engagement," has also been proposed, which involves a student constructively contributing to the flow of instructions \cite{reeve2011agency}.

The concept of engagement may differ based on the perspective from which it is being analyzed and the level of detail at which it is being studied; a concept referred to as "grain size" \cite{sinatra2015challenges}. When the grain size is considered at a micro level, engagement may refer to an individual's involvement in a specific task or learning activity. At a macro level, on the other hand, engagement may pertain to a group of students within a class or community. The National Survey of Student Engagement \cite{kuh2001assessing} is an example of a measure that is suited for evaluating engagement at the institutional level but may not be as effective in identifying correlations between an individual student's engagement and their learning experience.

There are various methods that have been employed to assess student engagement in traditional in-person learning settings, including self-reporting, observational scales, experience sampling, teacher rating, and interviews \cite{fredricks2012measurement}. Henrie et al. \cite{henrie2015measuring} conducted a review of the various self-reporting and observational (both qualitative and quantitative) scales that have been used to measure student engagement in technology-mediated learning environments and identified their strengths and limitations.

Virtual learning platforms often utilize video and audio mediums for both content delivery and communication between instructors and students. A variety of features, such as body pose, valence, arousal, and audio pitch, can be extracted from video and audio data \cite{karimah2022automatic,dewan2019engagement,abedi2021affect}. The extracted features can then be used to build AI models for engagement measurement. These features can also be learned through deep learning approaches \cite{karimah2022automatic,dewan2019engagement,abedi2021improving}. However, the extent of information captured is restricted due to limited modalities. Audio data, for example, cannot be used to learn or extract facial features. In theory, other types of sensors can also be employed, for example, electrocardiogram, electroencephalogram, and wearable devices to collect other physiological information (e.g., electrodermal activity, skin temperature, heart rate) \cite{bustos2022wearables}. However, in a real-world scenario, the use of many sensors in the educational environment and on the body of a student is impractical. Therefore, the key question to consider is whether these (extracted or learned) features from a sensing modality correspond or correlate to a measurement scale. Recent advancements in machine learning, especially deep learning, have allowed for the extraction of temporal affective and behavioral information from video and audio datasets for various tasks in the field of affective computing \cite{baltrusaitis2018openface,cao2019openpose}. However, in the context of SE, the existing virtual learning datasets used diverse observational scales for collecting ground-truth annotations (as discussed in Section \ref{sec:results}). Sometimes these scales are arbitrarily contrived, invented, or based on general knowledge rather than complete psychometric analysis. Therefore, there appears to be no direct concordance between the information extracted from the video, audio, or other sensing modalities and the measurement scales. In such cases, it is very hard to establish a clear interpretation between What we wanted to train an AI algorithm on and what actually the AI algorithm is trained on. In most of the existing AI-driven engagement measurement approaches \cite{karimah2022automatic,dewan2019engagement}, the focus was on building sophisticated AI models without as much emphasis on the correctness of annotations upon which they are trained. The outcomes of a successful AI model are as good as the quality of ground-truth annotations assigned to it \cite{liao2021deep,mehta2022three,abedi2021improving}. This further leads to questions about the validity of performance values reported by the existing AI-driven methods for SE measurement.

\subsection{‌Dimensions of Engagement Annotation}
\label{sec:dimensions}
D'Mello \cite{d2015influence} identified five dimensions of affect annotation for developing affect detection systems: \textit{sources}, \textit{data modality}, \textit{timing}, \textit{temporal resolution (timescale)}, and \textit{level of abstraction}. There are two key differences between affect detection and engagement measurement. Firstly, affect is only one component of engagement, which also includes behavioral and cognitive components, \cite{fredricks2004school}. Secondly, contrary to the affect "detection", an annotation for engagement "measurement" must not only identify engagement versus disengagement but also determine the "level" of engagement. Considering the differences between affect and engagement, we modified the five dimensions above and added two additional dimensions, \textit{combination} and \textit{quantification (scale)}, as described below. These seven dimensions of engagement annotation were used to analyze the SE definition and annotation used in the existing datasets.

\subsubsection{‌Sources}
\label{sec:sources}
The first dimension of engagement annotation, \textit{sources}, refers to the types and number of individuals performing the annotation \cite{d2015influence}. Observer-based annotation is the most common approach in the reviewed SE datasets (see Section \ref{sec:results}), which is categorized into expert (trained) observers and non-expert (or untrained) observers. An example of the expert observers, which is considerably high cost per observer, would be a group of educational psychology experts who are asked to annotate students' engagement in a dataset. Conversely, non-expert observers would be students without any prior training in psychology who are asked to annotate engagement in a dataset according to their perception of engagement. The annotation by non-expert observers is usually performed in crowdsourcing settings \cite{brabham2013crowdsourcing}. In crowdsourcing, a task is usually given over the internet to a less-specific and more-public-oriented group of observers. This approach can yield a large number of annotations in a short span of time. However, due to the lack of expertise in observers, it could lead to noisy annotations \cite{gupta2016daisee}.

Observer-based annotations do not interrupt the learning process of a student. However, they are time, and labor-consuming \cite{whitehill2014faces}, and can suffer from observation bias (such as \textit{seeing what one is looking for and missing what one is not} \cite{minner2010inquiry}). These measures are hard to scale but can "\textit{measure SE as it occurs}" \cite{henrie2015measuring}, which can be used to annotate various segments or transience of a student's engagement in a learning session. The observer-based measures are often used in conjunction with other measures for additional evidence rather than a stand-alone source of information \cite{o2016engagement}.

Annotations can also be performed by the student themselves, which is called self-report. Collecting concurrent self-report data at regular intervals can be disruptive and disengage individuals during their tasks \cite{o2016engagement}. Retrospective self-report also requires a student to reconstruct past states of engagement on a post-hoc basis, which may be biased. Different students may also differ in their own sense of what it means to be engaged. On the other hand, since SE also encompasses cognitive and emotional components, it is argued that self-report is the most valid measure to capture aspects of engagement that focus heavily on students' perception of their experience \cite{henrie2015measuring}.

\subsubsection{‌Data Modality}
\label{sec:dat_modality}
The \textit{data modality} dimension \cite{d2015influence} pertains to the information that is observed by observers for annotation, such as video, audio, computer screen recording, mouse cursor tracking data, or their combination.

\subsubsection{‌Timing}
\label{sec:timing}
Annotation \textit{timing} \cite{d2015influence} refers to the point at which the annotation takes place. This can take place in real-time, as when students are prompted to self-report their engagement through the use of pop-up windows during virtual learning sessions. In observer-based annotation, observers may be asked to watch the recorded (e.g., audio-visual) data of students and perform annotation in an offline manner \cite{aslan2017human} or in an online manner, e.g., live annotation of students' engagement in a learning session \cite{ocumpaugh2015baker}.

\subsubsection{‌Temporal resolution}
\label{sec:temporal_resolution}
\textit{Temporal resolution (timescale)} \cite{d2015influence} refers to whether the annotation is performed at frame-level (e.g., still images extracted from video frames), segment-level (e.g., pop-up window self-reports shown every 10 minutes in a session), or session-level (e.g., retrospective self-reports at the end of learning session). The majority of the existing engagement annotation datasets are recorded videos of students. Observers have either annotated single frames of videos, video segments of a predetermined length, or videos of the entire learning session. Some datasets have been annotated in an adaptive segment-level manner \cite{ocumpaugh2015baker,aslan2017human} in which the timescale (e.g., the length of video segments in a video dataset) is determined according to the changes in the engagement states of the students.

\subsubsection{‌Level of Abstraction}
\label{sec:level_of_abstraction}
Regarding the \textit{level of abstraction}, engagement can be annotated at a high level without considering the components of engagement, e.g., into two classes of engagement and disengagement. In a different setting, the affective, behavioral, and cognitive components of engagement are first separately annotated and then combined to result in a numerical value for engagement. In the field of affect annotation, Pomsta et al. \cite{porayska2013knowledge} have defined the above two settings as discrete response, and dimensional response, respectively. Each of the affective, behavioral, and cognitive components of engagement can also be annotated at different levels. To illustrate, behavioral engagement can be in two states of Off-Task and On-Task. The On-Task behavior itself can be in different categories of On-Task Conversation, On-Task Giving Answers, and so on \cite{ocumpaugh2015baker}.

\subsubsection{‌Combination}
\label{sec:combination}
An annotated dataset suitable for developing AI algorithms requires a numerical value or a class label for each sample in the dataset. The \textit{combination} dimension is concerned with how the annotated affective, behavioral, and cognitive components of engagement are combined to derive a numerical value or a class label for engagement. For instance, in the SE annotation protocol proposed by Aslan et al. \cite{aslan2017human}, the combination of the On-Task behavioral state and Highly-Motivated affective state results in the state of engagement (versus disengagement). Naibert et al. \cite{naibert2022development} proposed different architectures for the combination of affective, behavioral, and cognitive components of engagement collected through self-report questionnaires. It should be noted that the practice of combining the multiple components precludes examining distinctions among the components, and important information may be lost \cite{mcneish2022limitations}. In addition, a strategy for combination should take into account the correlation between the affective, behavioral, and cognitive states of students \cite{d2012dynamics}. As described for the level of abstraction, if one value is directly assigned to a specific level of engagement without considering its components, no combination is required.

\subsubsection{‌Quantification}
\label{sec:quantification}
The \textit{quantification (scale)} dimension refers to how engagement level is represented numerically, i.e., the type of the variable used for definition and annotation of engagement. The quantification of engagement level as a psychology term must ensure objectivity, precision, and rigor \cite{tafreshi2016quantification}. The engagement was quantified and annotated as a dichotomous (binary) variable having two states of engagement and disengagement. It was also quantified as a categorical variable in some datasets. Most datasets quantified engagement as an ordinal, or interval variable representing discrete, or continuous levels of engagement, respectively.

\section{Related Reviews}
\label{sec:comparison}
A few reviews have been published on automatic SE measurement, most of which focused on the AI methodologies and algorithms used for SE measurement. Dewan et al. \cite{dewan2019engagement} divided the existing SE measurement methods into three categories: automatic, semi-automatic, and manual, considering the methods’ dependencies on students’ participation. They identified the challenges involved in SE methods and briefly explored the available datasets and performance metrics for SE measurement techniques. Karimah and Hasegawa \cite{karimah2022automatic,karimah2021automatic} conducted a systematic review and studied available engagement definitions, datasets, and methods at a high level. After explaining the characteristics of the available datasets, they reviewed and explained different steps for engagement measurement as a supervised machine-learning problem. The authors covered pre-processing (including face detection, feature extraction, data augmentation, feature selection, dimensionality reduction, and imbalanced data handling), classic machine learning, deep learning, fine‑tuning and transfer learning and performance evaluation for engagement measurement methods. Salam et al. \cite{salam2022automatic} surveyed different aspects of context-driven engagement inference, entailing definition, engagement components and factors, publicly available datasets, ground truth assessment, features, and methods. The above aspects of engagement inference were studied in different settings, including human-human, human-computer, human-agent, and human-robot interaction. In the category of human-computer interaction, they covered a few SE in virtual learning datasets and measurement methods. Researchers in the area of educational psychology reviewed publications on the theory of student engagement, in general, \cite{wong2021student}, and in technology-mediated learning, \cite{henrie2015measuring,schindler2017computer,hu2017student}. The previous reviews in the area of computer science mainly were focused on machine-learning and deep-learning methodologies for engagement measurement \cite{dewan2019engagement,karimah2022automatic,karimah2021automatic,salam2022automatic}. Karimah and Hasegawa \cite{karimah2022automatic,karimah2021automatic} and Salam et al. \cite{salam2022automatic} briefly studied the existing SE datasets. However, the previous reviews did not emphasize the inconsistencies in the annotation of SE datasets and the difficulties associated with developing comparable AI models. We introduced the seven dimensions for engagement annotation (described in Section \ref{sec:dimensions}) to critically examine these inconsistencies. Additionally, we provide recommendations for appropriate SE definitions and annotation protocols in virtual learning environments.

\section{Methods}
\label{sec:methods}
IEEE Xplore, ACM Digital Library, SpringerLink, ScienceDirect, and Google Scholar were searched for English journals and conference publications published between 2010 and 2022. Different combinations of the following keywords were adopted: engagement measurement/detection/prediction/recognition/classification/regression, machine learning, deep learning, artificial intelligence, and dataset. Reviewers screened all studies in order to identify those in which a new dataset was proposed for developing AI models for automatic SE measurement. The studies introducing datasets containing single- or multi-modal data of individual students in online or offline computer-based virtual learning sessions were included. Correspondingly, the exclusion criteria are as follows: (i) the studies introducing SE datasets containing students in groups, (ii) the studies introducing SE datasets containing students in in-person classrooms, (iii) the studies introducing student datasets for general affect detection, such as basic affect state recognition or valence and arousal recognition, and not for engagement measurement, and (iv) the studies on developing AI models for SE measurement without introducing a new dataset. Although this review endeavored to provide a comprehensive overview of the literature, it should be noted that it was not conducted in a systematic review manner. The focus of this review was not on analyzing the use of AI techniques for SE measurement. Other reviews that delve into AI, machine learning, and deep learning \cite{dewan2019engagement,karimah2022automatic,karimah2021automatic,salam2022automatic} are available for interested readers, Section \ref{sec:comparison}.

The datasets were analyzed in terms of the definition and annotation of SE based on the seven dimensions of engagement annotation, described in Sections \ref{sec:dimensions}. In addition, the following seven baseline characteristics of the datasets were analyzed: computer-based lecture watching, reading activity, writing activity, or working with educational software; interactive or non-interactive activity; data collected in-the-wild or in-the-lab; the number of students; their sex; and age in the dataset; and the distribution of samples in different levels of engagement in the dataset.

\section{Results and Discussion}
\label{sec:results}
Thirty studies introducing new datasets for SE measurement in virtual learning were included. Tables 1, and 2 show the seven dimensions of engagement annotation in the existing datasets, and the baseline characteristics of the datasets, respectively. The inconsistencies in the seven dimensions of engagement in the dataset are analyzed as follows.

\subsection{‌Sources}
\label{sec:inconsistencies_in_sources}
The sources of annotation in the reviewed datasets were expert \cite{aslan2014learner,alyuz2021annotating,okur2017behavioral,alyuz2017unobtrusive,zaletelj2017predicting,bosch2016using,jeong2022evaluation}, non-expert \cite{whitehill2014faces,mohamad2020automatic,booth2017toward,kaur2018prediction,alkabbany2019measuring,bhardwaj2021application,binh2019detecting,gupta2022facial,verma2022multi}, crowdsourcers \cite{gupta2016daisee,kamath2016crowdsourced,delgado2021student,ruiz2022atl}, or through self-reported questionnaires \cite{vanneste2021computer,monkaresi2016automated,de2019engaged,chen2015video,hutt2019time,buono2022assessing,thomas2022automatic}. Some of the reviewed datasets also combined self-reports with observer-based annotation \cite{zheng2021estimation,altuwairqi2021new,ma2021hierarchical,bosch2016detecting,chen2016hybrid}. In most studies, non-expert or crowdsourcer observers were untrained students or freelancers. Booth et al. \cite{booth2017toward} pointed out that the observers did not receive any clarification or guidance regarding how to interpret the term engagement. In the event that the observers were unfamiliar with the concept they were annotating, they may have used their uninformed definition of engagement, which can lead to inaccurate annotations. The AI models built on such an annotation strategy could learn erroneous concepts. A specialized concept of SE needs to be annotated by either experts or people with training in the field. Otherwise, the validity of these labels may be under question. Gupta et al. \cite{gupta2016daisee} commented on noisy data after using a crowdsourcing platform for obtaining engagement annotations. In some other studies, noise filters were applied to labels \cite{kaur2018prediction}. Expert annotators and a validated SE definition and annotation protocol would prevent the need for such post-processing. The cost and time to annotate the data is a known challenge, but accurate data annotation is paramount to building generalizable AI models. Baker Rodrigo Ocumpaugh Monitoring Protocol (BROMP) \cite{ocumpaugh2015baker} and Human Expert Labeling Process (HELP) \cite{aslan2017human} (explained in Section \ref{sec:definitions}) are two engagement annotation protocols that are used in some of the studies, \cite{aslan2014learner,alyuz2021annotating,okur2017behavioral,alyuz2017unobtrusive,bosch2016using}, where training is provided to observers. Most of the questionnaires used in the reviewed datasets contained very few questions, sometimes including only one question \cite{monkaresi2016automated,booth2017toward}. A short questionnaire could indicate a flawed data collection process that could influence the value of reported metrics \cite{o2016engagement}.

In some of the reviewed datasets with multiple annotators, different inter-rater reliability or correlation metrics, e.g., Cronbach's alpha, Cohen's kappa, Fleiss' kappa, Krippendorff's alpha, and Pearson correlation, were used to evaluate the quality of the annotations produced by the annotators.

\subsection{Data Modality}
\label{sec:inconsistencies_in_data_modality}
A variety of types of information were used by observers for engagement annotation in the existing datasets, such as video \cite{whitehill2014faces,mohamad2020automatic,booth2017toward,aslan2014learner,alyuz2017unobtrusive,okur2017behavioral,alkabbany2019measuring,bhardwaj2021application,binh2019detecting,gupta2016daisee,kamath2016crowdsourced,delgado2021student,zheng2021estimation,altuwairqi2021new,ma2021hierarchical,bosch2016detecting,chen2016hybrid,buono2022assessing,jeong2022evaluation,verma2022multi}, image \cite{whitehill2014faces,mohamad2020automatic,gupta2022facial,ruiz2022atl}, audio \cite{alyuz2021annotating,alyuz2017unobtrusive}, screen capture \cite{aslan2014learner,alyuz2021annotating,alyuz2021annotating,okur2017behavioral,alyuz2017unobtrusive}, and mouse cursor tracking \cite{alyuz2021annotating,okur2017behavioral,alyuz2017unobtrusive}, or self-reports by students \cite{vanneste2021computer,monkaresi2016automated,de2019engaged,chen2015video,hutt2019time,buono2022assessing,thomas2022automatic}. The diversity of data modalities used for annotation across datasets may not pose a problem; however, the inconsistencies between the data modalities used for annotation and for developing AI models may cause problems. For instance, Chen et al. \cite{chen2015video} and Thomas et al. \cite{thomas2022automatic} used retrospective self-report questionnaires for annotation, but videos were used for training AI models. Retrospective self-reports are collected after the occurrence of engagement states. Therefore, as opposed to observer-based annotations with appropriate time resolutions, self-reports are not reliable reflections of the in-situ engagement states of students. Thus, it will be difficult to develop AI models on such data. It is important to investigate the capacity of different data modalities to represent different components of engagement. That is, to what extent each affective, behavioral, and cognitive component of engagement can be annotated based on which data modalities. For instance, Bosch \cite{bosch2016detecting} investigated the feasibility of measuring cognitive engagement using video data modality. According to Alyuz et al. \cite{alyuz2021annotating}, the distribution of affective and behavioral dimensions of engagement are different in students in different high school grades and in diverse ethnicities. The expression of emotions and affect also differs across genders \cite{brody2008gender}. This demographic information, such as sex, gender, age, ethnicity, students’ major, and the relevance of the virtual learning materials to students’ majors, should be taken into consideration in engagement data collection and annotation (see Tables 1 and 2).

\subsection{Timing}
\label{sec:inconsistencies_in_timing}
There were inconsistencies in the existing engagement datasets in terms of the timing of self-reports, most of which were annotated retrospectively \cite{monkaresi2016automated,de2019engaged,chen2015video,hutt2019time,buono2022assessing,thomas2022automatic}, and a few were based on concurrent self-reports \cite{vanneste2021computer,monkaresi2016automated,bosch2016detecting}. Moreover, Monkaresi et al. \cite{monkaresi2016automated} used a combination of both. Apart from only one dataset \cite{bosch2016using}, in which in-situ annotation was performed using BROMP protocol \cite{ocumpaugh2015baker}, all other observer-based annotations used retrospective annotation, which was performed using recorded data \cite{whitehill2014faces,mohamad2020automatic,booth2017toward,aslan2014learner,alyuz2017unobtrusive,alyuz2021annotating,okur2017behavioral,zaletelj2017predicting,bosch2016detecting,bosch2016using,khan2022unsupervised,alkabbany2019measuring,bhardwaj2021application,binh2019detecting,gupta2016daisee,kamath2016crowdsourced,delgado2021student,zheng2021estimation,altuwairqi2021new,ma2021hierarchical,bosch2016detecting,chen2016hybrid,gupta2022facial,jeong2022evaluation,ruiz2022atl}.

\subsection{Temporal Resolution}
\label{sec:inconsistencies_in_temporal_resolution}
D’Mello and Graesser \cite{d2012dynamics} have differentiated between mood states and affect states. While moods, e.g., depression, have been defined for an entire learning session (several minutes or a few hours), engagement, defined as an affect state, arises and decays at much faster timescales (a few seconds). According to the extensive experiments on the dynamics of affect states during learning \cite{d2012dynamics}, \cite{d2007monitoring}, and \cite{d2007dynamics}, there is an affect state transition approximately every 30 seconds, every 10-40 seconds, and every one minute. The temporal resolution of engagement annotation should be determined based on this affect dynamics (in different populations and in different learning situations). The existing datasets used inconsistent temporal resolutions for engagement annotation, starting from frame-level annotation \cite{whitehill2014faces,mohamad2020automatic,gupta2022facial,ruiz2022atl,alkabbany2019measuring,binh2019detecting,kamath2016crowdsourced,delgado2021student,ruiz2022atl}, segment-level annotation with segments of 1-second length to 30-minute length \cite{jeong2022evaluation,whitehill2014faces,booth2017toward,zaletelj2017predicting,kaur2018prediction,gupta2016daisee,zhang2019novel,ma2021hierarchical,bosch2016detecting}, and session-level annotation \cite{de2019engaged,thomas2022automatic}. Temporal resolution in some of the existing datasets is close to the timescales mentioned above, e.g., 10 seconds and 60 seconds in Whitehill et al. \cite{whitehill2014faces} and 10 seconds in Gupta et al. \cite{gupta2016daisee}. Whitehill et al. \cite{whitehill2014faces} have also reported higher inter-rater reliability for annotation with 10-second segments compared to 60-second segments.

None of the annotation protocols in the reviewed datasets with high temporal resolution indicated how to annotate when there is a transition between different levels of engagement. In the datasets with a relatively low temporal resolution, e.g., five minutes, more than one engagement state may occur in each timescale. Considering each data segment being annotated in the corresponding timescale as a multi-set (or bag of words) \cite{abedi2022detecting}, none of the existing annotation protocols determined how many engagement or disengagement states (words) must occur in the timescale to be annotated as engagement or disengagement. A plausible solution is to have an adaptive timescale as performed in BROMP \cite{ocumpaugh2015baker} and HELP \cite{aslan2017human} annotation protocols (explained in Section \ref{sec:definitions}).

Inconsistencies in temporal resolution make it difficult to develop and evaluate AI models across datasets. For instance, a sequential machine-learning model with a specific architecture \cite{abedi2021affect,abedi2022detecting,liao2021deep} is not capable of simultaneously handling video segments of lengths 10 seconds in the dataset presented in \cite{gupta2016daisee}, 100 seconds in \cite{thomas2022automatic}, and 5 minutes in \cite{kaur2018prediction}.

\subsection{Level of Abstraction}
\label{sec:inconsistencies_in_the_level_of_abstraction}
Even though engagement is a multi-component state \cite{fredricks2004school,sinatra2015challenges}, in most of the reviewed datasets, it was defined as a single-component state without clarification of its components \cite{buono2022assessing,jeong2022evaluation,thomas2022automatic,vanneste2021computer,monkaresi2016automated,booth2017toward,aslan2014learner,zaletelj2017predicting,bhardwaj2021application,binh2019detecting,kamath2016crowdsourced,ma2021hierarchical,ruiz2022atl}. Some authors annotated engagement based on only one of its components, e.g., affective \cite{gupta2022facial,chen2015video,hutt2019time,gupta2016daisee,altuwairqi2021new}, behavioral \cite{okur2017behavioral,alyuz2017unobtrusive,kaur2018prediction,delgado2021student}, or cognitive engagement \cite{bosch2016detecting}. Some works annotated engagement as a multi-component state \cite{whitehill2014faces,mohamad2020automatic,alyuz2021annotating,bosch2016using,alkabbany2019measuring}. The existing datasets containing single or multi-modal data did not provide any rationale for considering only one or more components of engagement.

\subsection{Combination}
\label{sec:inconsistencies_in_combination}
In the datasets in which engagement was defined and annotated as a multi-component state, the components of engagement were combined to generate one numerical value for engagement \cite{mohamad2020automatic,alyuz2021annotating}, as in HELP protocol \cite{aslan2017human}, in which various combinations of affective and behavioral components resulted in a dichotomous engagement state. In some other works, one component of engagement was taken into consideration as a prerequisite for other components of engagement. For instance, in Alkabbany et al. \cite{alkabbany2019measuring}, the presence of behavioral engagement (and the absence of affective engagement) corresponded to lower levels of engagement. Then, the presence of both affective and behavioral engagements corresponded to higher levels of engagement. In some other works, such as the datasets in which BROMP \cite{ocumpaugh2015baker} was used for annotation, engagement was annotated as an affective state, and behavioral states were annotated separately \cite{bosch2016using}. They did not combine the affective and behavioral dimensions. In a totally different engagement annotation method, Bosch \cite{bosch2016detecting}, the occurrence of mind-wandering was considered as cognitive engagement.

\subsection{Quantification}
\label{sec:inconsistencies_in_quantification}
A major issue in the reviewed datasets is the inconsistency in the use of scales to measure SE. A few researchers, e.g., Vanneste et al. \cite{vanneste2021computer}, Zaletelj and Ko{\v{s}}ir \cite{zaletelj2017predicting}, have stressed the fact that there is no "gold standard" for measuring engagement. There are different quantification methods used in different datasets to represent different levels of engagement, and these methods range from the use of two points to the use of six points, as well as the use of continuous values. Correspondingly, engagement was considered a dichotomous (binary) variable \cite{monkaresi2016automated,mohamad2020automatic,binh2019detecting,bosch2016detecting,chen2016hybrid,jeong2022evaluation}, an ordinal variable \cite{chen2015video,hutt2019time,whitehill2014faces,zaletelj2017predicting,kaur2018prediction,alkabbany2019measuring,bhardwaj2021application,gupta2016daisee,kamath2016crowdsourced,zheng2021estimation,altuwairqi2021new,thomas2022automatic}, or an interval variable \cite{vanneste2021computer,booth2017toward,ma2021hierarchical,gupta2022facial,buono2022assessing}. Some datasets defined engagement as a categorical variable \cite{aslan2014learner,alyuz2021annotating,okur2017behavioral,alyuz2017unobtrusive,delgado2021student,ruiz2022atl}, e.g., three categories of Engaged, Not-engaged, and Unknown in \cite{aslan2014learner}. Moving forward, this inconsistency can be a major constraining factor for progress in the field. In a simplified sense, the concept of an object, "X" must be consistently annotated as "X" based on a commonly accepted measurement instrument across multiple data sources. Otherwise, supervised machine-learning models may not be able to learn that concept effectively. Corresponding to the different engagement scales in the existing datasets, different types of supervised machine-learning models were trained to solve binary or multi-class classification or regression problems \cite{karimah2022automatic,dewan2019engagement}. Due to this scale inconsistency, it is infeasible to use a machine-learning model trained on one dataset to make engagement inferences on another dataset annotated with a different engagement scale. Moreover, the performance of machine-learning models trained on different datasets with different engagement scales (e.g., a binary classification model with a regression model) cannot be compared.

\subsection{Miscellanies}
\label{sec:miscellanies}
\subsubsection{Publicly Available Datasets}
\label{sec:publicly_available_satasets}
A limited number of the reviewed datasets were available publicly, including Dataset for Affective States in E-Environments (DAiSEE) \cite{gupta2016daisee}, Emotion Recognition in the Wild-Engagement prediction in the Wild (EmotiW-EW) \cite{kaur2018prediction}, and Affect Transfer Learning for Behavior Prediction (ATL-BP) Student Engagement Dataset \cite {delgado2021student,ruiz2022atl}. Few researchers have analyzed and pointed out problems with the annotations in the public datasets. Abedi and Khan \cite{abedi2021improving} indicated the annotation issues in the DAiSEE dataset as a misleading factor in training temporal and non-temporal deep-learning models. Additionally, Liao et al. \cite{liao2021deep} and Mehta et al. \cite{mehta2022three} discussed the annotation problems in the DAiSEE dataset by providing examples of videos and annotations of one student in different levels of engagement. Specifically, Liao et al. \cite{liao2021deep} criticized the use of discrete labels when annotating SE levels in videos and proposed the use of continuous values instead.

\subsubsection{Characteristics of Virtual Learning Environment}
\label{sec:characteristics_of_the_virtual_learning_environment}
The characteristics of the virtual learning environment in which the engagement annotation definition and protocol are designed to be applied are another important missing factor in the existing datasets. The existing datasets did not provide a rationale or justification for using a particular SE definition with respect to the characteristics of the virtual learning setting in the dataset. For example, it is important to determine whether the virtual learning environment is interactive or non-interactive. An interactive setting involves more interaction between the student and the computer (e.g., mouse cursor movements) than a non-interactive setting in which the student merely watches a recorded or online video. It is important to consider whether it is: (i) an online course with live communication between students and instructor, (ii) a recorded video of the instructor being viewed offline on a computer, (iii) a writing task on the computer screen, or (iv) a writing task on a piece of paper. As the affective, behavioral, and cognitive components of engagement are different in the above exemplary settings, the characteristics of the virtual learning setting must be considered during the design of engagement annotations.

\subsubsection{Imbalanced Distribution}
\label{sec:imbalanced_distribution}
The distribution of samples in different levels of engagement in the existing datasets is presented in Table 2. It can be observed that the number of samples in disengagement or low levels of engagement is typically much lower than the number of samples in higher levels of engagement in almost all the existing datasets. The highly imbalanced data distribution in these datasets must be considered when developing AI models.

\section{Student Engagement Definitions and Protocols in Other Settings}
\label{sec:definitions}
The definitions of SE presented in Section \ref{sec:student_engagement_annotation} and research in education and psychology have led to the definition and design of several SE scales used in various settings. In this section, some of the existing SE definitions and protocols in settings other than virtual learning that have the potential to be used in measuring SE in virtual learning are discussed.

BROMP \cite{ocumpaugh2015baker} is an observation protocol for in vivo annotation of students’ affective and behavioral states. In the BROMP platform, observers (\textit{sources}) are trained and tested on the BROMP annotation protocol and achieve sufficient inter-rater agreement, Cohen’s Kappa > 0.6, in their observations and get a BROMP certification before participating in the annotation. Students in an in-person classroom working with educational software on computers are observed by the observer in-person by a side glance to make a holistic judgment of a student’s state based on facial expressions, speech, body posture, gestures, and the student’s interaction with the educational software (\textit{data modality}). Observation is performed in a round-robin manner, observing and annotating one student and moving to the next. The frequency of observations per student varied between class periods depending on the number of students in the class (\textit{timing}). Each student is observed for 20 seconds or until a visible state is detected (\textit{temporal resolution}). The annotation is inserted in a mobile application. In the BROMP protocol, the affective and behavioral states of students are annotated separately (\textit{combination}). Various affect states are included in the BROMP protocol; some commonly used are Boredom, Confusion, Delight, Engaged Concentration, Frustration, and Surprise. The main behavioral states are On-Task and Off-Task (\textit{level of abstraction} and \textit{quantification}).

Aslan et al. \cite{aslan2017human} stated unaddressed challenges in BROMP \cite{ocumpaugh2015baker} as follows: (1) limited chance for revision as annotation is performed in vivo, (2) difficulty of making a decision about a student's state in real-time, (3) fragmented annotation and disregarding state change in students due to the round-robin technique, (4) limited labels for model training, and (5) inevitable observer effect due to the presence of the observer in the learning settings. To address these challenges, they developed the HELP annotation process. HELP has a systematic process of training and evaluation for observers (\textit{source}). It contains an annotation software containing the recorded video, audio, screen capture of students' computers and learning material contextual data, and demographic information of students (\textit{data modality}). The \textit{timing} is post-facto as observers watch the recorded data of students retrospectively. \textit{Temporal resolution} is similar to BROMP, after observing the first state change of the student. The annotation of the affect and behavioral states are separate. The discrete affective states are Satisfied, Bored, Confused, and the behavioral states are On-Task and Off-Task (\textit{level of abstraction} and \textit{scale}). Inspired by Woolf et al. \cite{woolf2009affect}, different combinations of affect and behavioral states result in the dichotomous state of Engaged versus Not-engaged (\textit{combination}). Some studies have used BROMP \cite{bosch2016using} and HELP \cite{alyuz2021annotating,okur2017behavioral,alyuz2017unobtrusive} for engagement annotation. Aslan et al. \cite{aslan2017human} failed to demonstrate how addressing the fifth challenge in BROMP regarding the "inevitable observer effect" resulted in a better annotation. It also needs to be investigated which technique is optimal, observing one student continuously in HELP or using the round-robin technique in BROMP.

Altuwairqi et al. \cite{altuwairqi2021new} presented an affective model (not an annotation protocol) for engagement, in which different areas of the circumplex model of affect \cite{russell1980circumplex}, corresponding to different values of valence and arousal, are defined as five ordinal levels of engagement: Disengagement, Low, Medium, High, and Strong Engagement. In combination with self-reports, this affective model of engagement was used for video-based engagement annotation by Altuwairqi et al. \cite{altuwairqi2021student}.

Deng et al. \cite{deng2020learner} developed the Massive Open Online Course (MOOC) Engagement Scale (MES). The scale is a 12-item questionnaire comprised of four components, behavioural engagement, emotional engagement, cognitive engagement, and social engagement. The students are asked to fill out the questionnaire at the end of MOOC using a 6-point Likert scale for each item. Therefore, the timing and temporal resolution of MES are after the course, and the entire course semester, respectively. The items were also designed based on the timing, e.g., "I set aside a regular time each week to work on the MOOC". The Online Student Engagement scale (OSE) is a 19-item questionnaire developed by Dixson \cite{dixson2015measuring}. Students report on a 5-point Likert scale how well each behavior, thought, or feeling was characteristic of them or their behavior. Examples of questions in OSE are "Making sure to study on a regular basis" and "Doing well on the tests/quizzes".

As is the case with the inconsistent SE annotation scales in the existing virtual learning datasets, the annotation protocols discussed in this section were also inconsistent with respect to the seven dimensions of engagement annotation. These annotation protocols were not specifically designed for the purpose of SE annotation in virtual learning settings and to develop AI models. However, with appropriate modifications (considering the seven dimensions of engagement annotation), these protocols have the potential to be utilized for this purpose, \cite{aslan2014learner,alyuz2021annotating,okur2017behavioral,alyuz2017unobtrusive,bosch2016using}.

\section{Conclusions}
\label{conclusions}
In this critical review, we examined the existing SE virtual learning datasets and highlighted inconsistencies in terms of engagement definitions and annotations. Our analysis was based on the seven dimensions of engagement annotation: sources (observers), data modality, timing, temporal resolution, level of abstraction, combination, and quantification (scale). We discussed to what extent different dimensions of engagement annotation in the existing datasets are in accordance with the definition of SE in educational psychology. We explained how the inconsistencies are problematic in developing AI models for automatic SE measurement. We discussed some of the existing SE definitions and protocols in settings other than virtual learning that can be used in measuring SE in virtual learning. We appreciate the previous work by researchers who collected these datasets in order to contribute to progress in the field. However, we also raised doubts about the comparability of labels of different datasets and the generalizations of AI models across these datasets. We strongly recommend in future studies that both the observer-based and self-reporting SE annotations should be used in tandem to provide better evidence of SE in virtual learning. Consistent approaches for observational and self-reporting measurement of engagement should be developed to make progress in the field of developing AI models for SE measurement.

\begin{landscape}
\includepdf[landscape=true,pages=1]{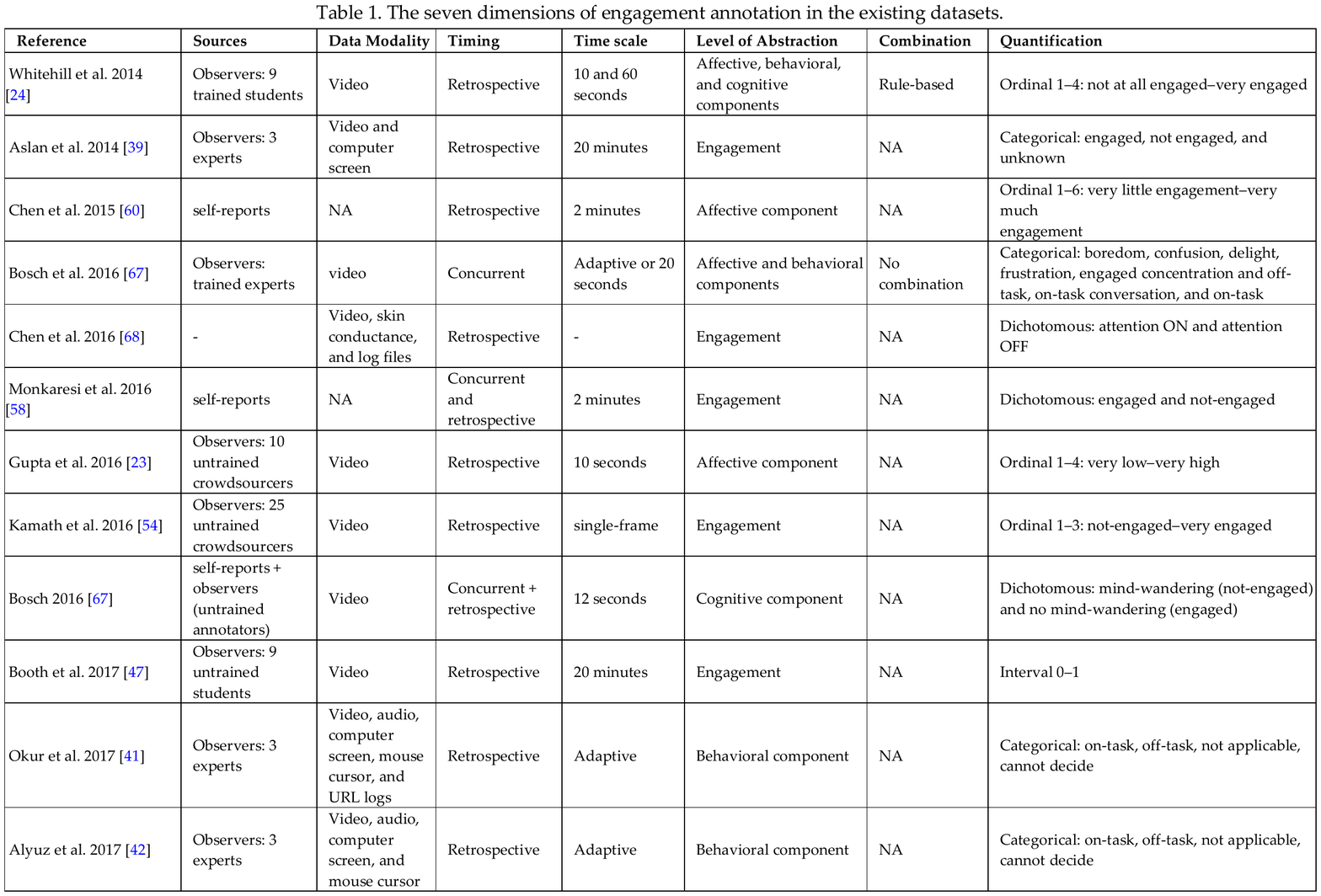}
\includepdf[landscape=true,pages=2]{that.pdf}
\end{landscape}

\begin{landscape}
\includepdf[landscape=true,pages=1]{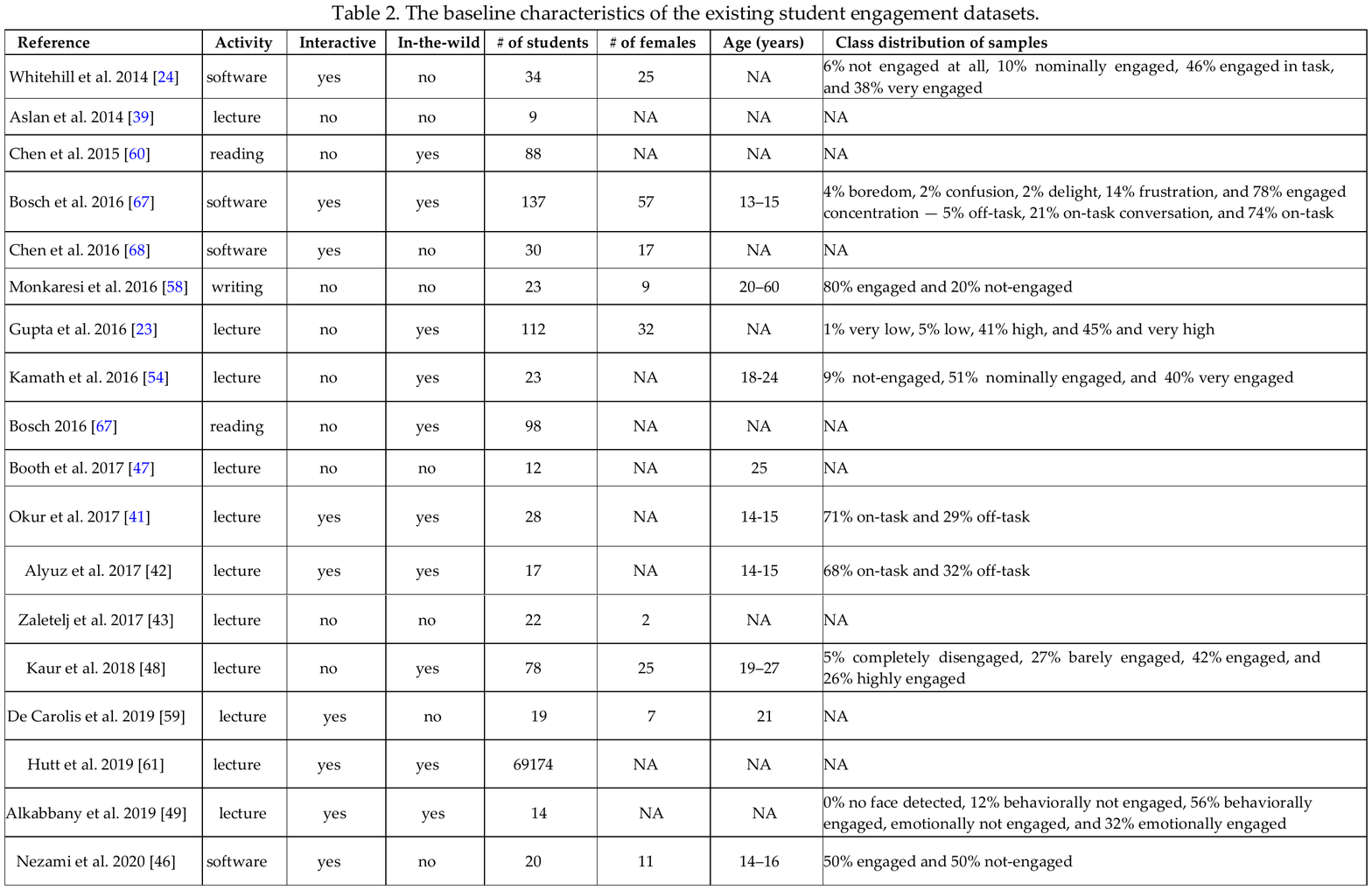}
\includepdf[landscape=true,pages=2]{this.pdf}
\end{landscape}

\textbf{Data availability}\\
No dataset was used.
\\\\
\textbf{Conflict of Interest}\\
The authors declare that they have no conflict of interest.

\bibliography{sn-bibliography}
\end{document}